\documentclass{jpp}
\usepackage{graphicx}
\usepackage{epstopdf, epsfig}
\usepackage{hyperref}
\usepackage{journals}
\usepackage{amsmath}
\usepackage{color}

\shorttitle{Simulating the global accretion disc dynamo}
\shortauthor{Nixon, Pringle \& Pringle}

\title{On the role of numerical diffusivity in MHD simulations of global accretion disc dynamos}

\author{C.~J.~Nixon\aff{1}
  \corresp{\email{C.J.Nixon@leeds.ac.uk}},
  C.~C.~T.~Pringle\aff{2}
   \and J.~E.~Pringle\aff{3}}

\affiliation{\aff{1}School of Physics and Astronomy, Sir William Henry Bragg Building, Woodhouse Ln., University of Leeds, Leeds LS2 9JT, UK
\aff{2}Centre for Fluid and Complex Systems, Coventry University, Coventry, CV1 5FB, UK
\aff{3}Institute of Astronomy, University of Cambridge, Madingley Road, Cambridge, CB3 0HA, UK}

\begin{document}

\maketitle

\begin{abstract}
Observations, mainly of outbursts in dwarf novae, imply that the anomalous viscosity in highly ionized accretion discs is magnetic in origin, and requires that the plasma $\beta \sim 1$. Until now most simulations of the magnetic dynamo in accretion discs have used a local approximation (known as the shearing box). While these simulations demonstrate the possibility of a self-sustaining dynamo, the magnetic activity generated in these models saturates at $\beta \gg 1$. This long-standing discrepancy has previously been attributed to the local approximation itself. There have been recent attempts at simulating magnetic activity in global accretion discs with parameters relevant to the dwarf novae. These too find values of $\beta \gg 1$. We speculate that the tension between these models and the observations may be caused by numerical magnetic diffusivity. As a pedagogical example, we present exact time-dependent solutions for the evolution of weak magnetic fields in an incompressible fluid subject to linear shear and magnetic diffusivity. We find that the maximum factor by which the initial magnetic energy can be increased depends on the magnetic Reynolds number as ${\mathcal R}_{\rm m}^{2/3}$. We estimate that current global numerical simulations of dwarf nova discs have numerical magnetic Reynolds numbers around 6 orders of magnitude less than the physical value found in dwarf nova discs of ${\mathcal R}_{\rm m} \sim 10^{10}$. We suggest that, given the current limitations on computing power, expecting to be able to compute realistic dynamo action in observable accretion discs using numerical MHD is, for the time being, a step too far.
\end{abstract}

\section{Introduction}
Fluid-based magnetic dynamos can be thought of as coming in two flavours:  small-scale and large-scale. Both types can be found in different astrophysical contexts \citep[see the reviews by][]{Brandenburg:2005aa,Rincon:2019aa,Schekochihin:2022aa}. Small-scale dynamos tend to produce magnetic fields that are correlated on the length-scale, $l_0$, (or smaller) of the driving turbulence, such as is seen in models of the internal dynamics of molecular clouds \citep[e.g.\@][]{Padoan:2014aa,Federrath:2016aa}, or in cosmological simulations \citep[e.g.\@][]{Martin-Alvarez:2021aa}. Large-scale dynamos typically comprise small-scale turbulence (scale $\l_0$) set within a large scale shear flow (scale $L \gg l_0$), and show large-scale spatial coherence. Obvious examples of these are the dynamo responsible for the solar cycle, and the dynamo thought to be responsible for driving the accretion flow within accretion discs. In the former, the large scale flow is the differential rotation of the sun, and the turbulence is driven by convective motions. In the latter, the large scale flow is the Keplerian differential flow of the disc, and the small scale turbulence is due to the magneto-rotational instability \citep[MRI;][]{Brandenburg:1995aa,Balbus:1998aa}, perhaps enhanced by magnetic buoyancy \citep{Tout:1992aa,Johansen:2008aa}.

Due to the complexity of the resulting dynamics, particularly in the nonlinear, turbulent phase, it is routine to resort to numerical simulations to gain an understanding of the flows generated from such dynamo action. For many years it was not straightforward to resolve the lengthscales associated with the growth of the MRI in global accretion disc simulations, and it is therefore necessary to employ a local, or ``shearing box'', approach to study the dynamics in a simplified framework with less computational power required to achieve high resolution. However, the shearing box approach does not reach the level of dynamo activity that is required to explain the observations. \cite{King:2007aa} note that the shearing box approach necessarily restricts the available modes that can be produced in the flow, and in particular low-$m$ modes (with $m \ne 0$) are not captured \citep[see also, for example, the discussion in][]{Parkin:2013aa}. There also remain questions regarding the convergence of such models when they are extended to very high resolution \citep{Bodo:2014aa}. We do not pursue such arguments here.

The discrepancy in the strength of the dynamo activity in shearing box models and observed accretion discs motivates the development of global MHD models of accretion discs. However, the computational demands of global models mean that the spatial resolution, typically measured as a fraction of the local disc scale height or the wavelengths of the fastest growing modes, may not be sufficient. In particular insufficient resolution may imply strong levels of numerical magnetic diffusivity compared with the physical diffusivity expected in the simulated systems. If the numerical diffusivity is far greater than the expected physical diffusivity, then it seems reasonable to expect that this will have implications for the types (strengths and geometries) of fields that are produced in the simulations compared to those produced in real systems \citep[cf.][]{Tobias:2013aa}.

In Section~\ref{discs} we briefly describe the relevant background to accretion disc physics and the relevant observed properties of such discs. In Section~\ref{discdynamoobs} we present what is known about the properties of the disc dynamo as can be deduced from observations of the discs in dwarf novae (a particularly well-studied subclass of cataclysmic variable stars in which a white dwarf accretes from a companion star). The discs in the outburst state of these objects are those for which we have the best understanding of the properties of the so-called anomalous viscosity. In these objects, if this viscosity is magneto-hydrodynamic in origin, then the observations imply that the magnetic fields are strong (plasma $\beta \sim 1$). In Section~\ref{discnum} we discuss the numerical simulations of the MHD properties of accretion discs. We note that in the usual shearing box approximation, it has long been known that the predicted magnetic field strengths are too low ($\beta \gg 1$) to satisfy the observations. It is possible that here the problem lies with the shearing box approximation itself, although we must remark that there is no evidence that this is the case. We therefore consider some recent numerical simulations of global accretion discs, but we note that in these too, the field strengths that are found are similarly small. In Section~\ref{missing}, we speculate that for the global disc simulations the numerical magnetic diffusivity may be too large to allow the required growth of the field. To illustrate this, in Section~\ref{example} we present calculations of the evolution of magnetic fields embedded in a fluid subject to an inexorable shear. We vary the level of diffusivity to illuminate the effect of it on the maximum magnetic field growth that is achievable. We highlight the limitations if excessive diffusion is included either explicitly or through numerical effects. Finally, we discuss our results in the context of some of the global simulations presented in the literature and draw conclusions in Section~\ref{discussion}.

\section{Accretion Discs}
\label{discs}
The theory of accretion discs is set out by \citet[][see also \citealt{Pringle:1981aa,Frank:2002aa}]{Shakura:1973aa}. The basic disc flow, in cylindrical polar $(R, \phi, z)$ coordinates, is in the azimuthal direction, 
\begin{equation}
u_\phi = \sqrt{GM/R},
\end{equation}
where $M$ is the central gravitating mass. The disc thickness or scaleheight, $H$, in the $z$-direction is given by 
\begin{equation}
H/R \approx c_s/u_\phi,
\end{equation}
where $c_s$ is the local sound speed. For the usual thin disc approximation we require that $H/R \ll 1$. Thus the azimuthal motion is supersonic, with Mach number $\sim R/H$. Inflow (i.e. accretion) through the disc requires the action of a so-called ``anomalous viscosity'' which taps the azimuthal ($R \phi$) shear and transfers angular momentum outwards and mass inwards. The viscosity is generally assumed to be caused by small-scale, $l_0 \le H$, magneto-hydrodynamic turbulence within the disc, and it is the maintenance of this turbulence that requires dynamo action. \cite{Shakura:1973aa} introduced  a parameter $\alpha$ which is a dimensionless measure of the  size of the anomalous viscosity -- essentially a dimensionless measure of the $z$-averaged $R\phi$-stress. Thus the ($z$-averaged) effective kinematic viscosity of the disc may be written as 
\begin{equation}
\nu \approx \alpha c_s H \approx \alpha H^2 \Omega,
\end{equation}
where 
\begin{equation}
\Omega = u_\phi/R = \sqrt{GM/R^3}
\end{equation}
is the angular velocity at radius $R$. For MHD turbulence, they note that 
\begin{equation}
\alpha \approx \langle B_R B_\phi \rangle / \rho c_s^2 \label{magstress}
\end{equation}
(in appropriate units), where $\rho$ is the disc density.  They also introduced physical arguments as to why we might expect $\alpha \le 1$ \citep[see also the discussion in][]{Martin:2019aa}. In terms of $\alpha$, the radial accretion flow velocity is 
\begin{equation}
u_R \sim - \alpha (H/R) c_s
\end{equation}
(where the minus sign indicates inward flow), and is subsonic.

\section{The disc dynamo -- observable  properties}
\label{discdynamoobs}
The stars for which we have the most comprehensive understanding of the properties (thermal, magnetic) of the dynamo fluid, and for which we have the best handle on the global properties of the dynamo itself are the subclass of the cataclysmic variable stars, known as dwarf novae. The cataclysmic variables are binary stars, consisting of a low-mass, solar-type star which is losing its outer layers to its companion. The companion is a more massive, but more compact star, about the size of the Earth, known as a white dwarf \citep{Warner:1995aa}.  Because of angular momentum considerations, the flow takes the form of an accretion disc \citep[see, for example,][]{Pringle:1985aa}. The dwarf novae show two states: (i) a bright outburst state in which the mass accretion rate onto the white dwarf is high and the disc is highly ionized, with low magnetic diffusivity, and (ii) a dim quiescent state in which the accretion rate is low, the disc ionization is low and the magnetic diffusivity relatively high. 

\subsection{Hot, highly-ionized discs}
The evolution of the surface density of an accretion disc is described by a diffusion equation, with the diffusion parameter proportional to the disc kinematic viscosity, $\nu$, that is, proportional to the Shakura-Sunyaev parameter $\alpha$ \citep[see, for example,][]{Lynden-Bell:1974aa,Pringle:1981aa}. During the decay from a dwarf nova outburst, the mass drains from the disc on to the central white dwarf. The timescale for this decay depends directly on the value of $\alpha$. The decay timescale of the outburst allows for a measurement of $\alpha$ in the hot state from modeling the outburst light curve. The disc size is known from the properties of the system and the disc temperature is obtained from the spectra. A simple calculation by \cite{Bath:1981aa} suggested that $\alpha \approx 1$. Since then there has been extensive and more detailed modelling of dwarf nova outburst behaviour, and all models point to relatively large values of $\alpha$. The models imply that $\alpha \approx 0.2 - 0.3$ \citep[e.g.][]{Pringle:1986aa,Smak:1998aa,Smak:1999aa,Buat-Menard:2001aa,Cannizzo:2001ab,Cannizzo:2001aa,Schreiber:2003aa,Schreiber:2004aa,Balman:2012aa,Kotko:2012aa,Coleman:2016aa}.

It should be noted that these measurements are in line with the values of $\alpha$ deduced from time-dependent disc behaviour in other systems with highly ionized accretion discs, for example the soft X-ray transients and the Be stars \citep[see the discussion in][]{Martin:2019aa}.

If, as we assume here, the dominant stresses that give rise to $\alpha$ are magnetic, this implies (see equation~\ref{magstress}) that the magnetic pressure in the disc is comparable to the gas pressure. Indeed, formally, since for the MHD dynamo driven by ($R \phi$) shear, we expect that $\langle B_\phi^2 \rangle \gg \langle B_R^2 \rangle$, it is evident that we require $\langle B_\phi^2 \rangle/ \rho c_s^2 \gtrsim 1$. In other words, the observations seem to imply that the $\beta$ parameter of the disc plasma is such that $\beta \lesssim 1$.

From the observed disc properties, we may estimate the physical value of the magnetic diffusivity, $\eta$, and hence the magnetic Reynolds number defined as  
\begin{equation}
{\mathcal R}_{\rm m} = c_s H /\eta,
\end{equation}
at a typical point in the plane of a dwarf nova accretion disc in outburst. 

We take the central star to have a mass $M = 1 M_\odot = 2 \times 10^{33}$\,g, and consider a typical radius in the disc, $R = 10^{10}$\,cm. For a highly ionized disc, in the bright state of a dwarf nova, we take a typical accretion rate of ${\dot M} = 10^{18}$\,g/s, and we take the dimensionless viscosity parameter to be $\alpha = 0.3$ in line with observations.

We evaluate $\Omega_0$ as
\begin{equation}
\Omega_0 = \left( \frac{GM}{R^3} \right)^{1/2} = 1.2 \times 10^{-2} \,{\rm radian/s}\,.
\end{equation}

For the magnetic diffusivity we assume the usual Spitzer value of $\eta = 3.5 \times 10^{12}  \, T^{-3/2}$ cm$^2$/s \citep{Potter:2014aa,Spitzer:1962aa}. From \cite{Frank:2002aa} we find that the temperature in the plane of the disc is $T_{\rm c} = 7.1 \times 10^4$ K. Thus we have that $\eta = 1.9 \times 10^5$ cm$^2$/s. Similarly we find that the disc scaleheight is given by $H = 3.8 \times 10^8$ cm, so that the disc opening angle is $H/R = 0.038$.

From these we are able to estimate a typical magnetic Reynolds number as
\begin{equation}
{\mathcal R}_{\rm m} = 9.1 \times 10^{9}\,.
\end{equation}

In summary, we note that, in these discs, whatever the origin of the turbulent behaviour within the disc that gives rise to the observed effective viscosity, whether it is purely hydrodynamic, or (as is generally believed) magneto-hydrodynamic, the mechanism that produces it is able to drive the fluid motions only up to, or close to, the sound speed. The fact that $\alpha$ is always found to be close to this limit (for these discs) implies that whatever instability might give rise to the driving mechanism for the turbulence,  the turbulent velocities grow until they become trans-sonic. 

Thus, in agreement with the original conjecture of \cite{Shakura:1973aa}, saturation of the turbulence is achieved once the motions become trans-sonic. This must be the result of the fact that once the motions approach the sound speed, the nature of the turbulence changes in a fundamental fashion. In line with the ideas of \cite{Shakura:1973aa,Shakura:1976aa}, it is evident that the change in the nature of the turbulence might occur for one, or both, of two physical reasons. First, in the case of hydrodynamic turbulence, as the turbulence becomes trans-sonic shocks begin to dominate the dissipative process. Second, in the case of MHD turbulence, once the Alfv\'en speed  approaches the sound speed (i.e. once $\beta \approx 1$),  not only do the turbulent velocities become trans-sonic, but, in addition, the timescale for the Parker instability (leading to loss of magnetic flux from the disc) becomes comparable with the shearing timescale (growth timescale for magnetic flux) $\approx 1/\Omega$ \citep[cf.][]{Tout:1992aa}.

The corollary of this basic finding is that numerical simulations of disc turbulence (for highly ionized discs) which do not find that the strength of the turbulence grows until limited by the sound speed (and which therefore do not find the large values of $\alpha$ implied by the observational data) cannot provide an adequate description of observed accretion disc dynamos. It seems likely that such models must be missing some fundamental physics \citep{King:2007aa}.

\subsection{Cool discs}
The value of $\alpha$ in the low state dwarf nova accretion disc is difficult to determine, as the disc in that state shows little in the way of time-evolution, and what time-evolution there is appears to be dominated by the continued addition of mass to the disc from the companion star. In addition, estimates of $\alpha$ in the quiescent disc require modelling of the complete outburst cycle. However, all models of dwarf nova outburst cycles seem to require that in the quiescent disc the value of $\alpha$ is less than that found in the outburst disc by at least a factor of $\approx 10$ \citep{Meyer:1983aa,Pringle:1986aa,Cannizzo:1993aa,Cannizzo:2001aa,Coleman:2016aa}.

These findings are in line with the idea (suggested for the dwarf novae quiescent discs by \citealt{Gammie:1998aa}) that the driving force for MHD disc turbulence, namely the MRI, is suppressed once the magnetic diffusivity becomes too large. A number of (local, shearing box) simulations suggest that the MRI becomes inoperative once ${\mathcal R}_{\rm m} \lesssim 10^3 - 10^4$ \citep{Hawley:1996aa,Stone:1996aa,Fleming:2000aa,Davis:2010aa}.  These results strengthen the argument that the anomalous viscosity (at least in dwarf novae) is an MHD effect \citep{Martin:2019aa}.

\section{The disc dynamo - numerical simulations}
\label{discnum}

\subsection{Local simulations -- the shearing box}
\label{local}
Most simulations of disc dynamos make use of the shearing box approximation \citep{Hawley:1995aa}.  Here the computational domain is a Cartesian box, of size $\sim H \ll R$ co-moving with the fluid at some fixed radius, $R_0$, where the angular velocity is $\Omega_0$. Thus $(R, \phi, z) \rightarrow (x = R - R_0, y = \phi - \Omega_0 t, z)$. The base flow in the box is a linear shear $u_y = U^\prime x$, where in a Keplerian disc $U^\prime = \frac{3}{2} \Omega_0$, and the box rotates with angular velocity $\Omega_0$. In the simulations the disc may, or may not, be stratified in the $z-$direction.

Typically, the simulations start with a small initial field. For example, \cite{Brandenburg:1995aa} and \cite{Hawley:1996aa} take an initial poloidal field, with zero net flux, of the form $B_z \propto \sin k x$. The early results are summarized in a review by \cite{Balbus:1998aa}. The most important finding was that with small initial seed fields, a steady, turbulent MHD disc dynamo could occur. However, for seed fields which had no externally imposed net field (i.e. for those simulations relevant to astrophysical discs) the magnitude of the $R \phi$ magnetic stress was around $\alpha \sim 0.01$, about an order of magnitude smaller than that required by observations. \cite{Balbus:1998aa} conceded that while the dynamo saturation with $\alpha \approx 1$ postulated by \cite{Shakura:1973aa} was an attractive physical possibility, this was not what was found in the simulations. They concluded: ``It appears likely, therefore, that there is a dynamo regime that is characterized by unstable growth continuously balancing dissipation scale losses. This leads to subthermal magnetic fields and a dimensionless stress tensor  of order $\alpha \approx 0.01$ . Whether there are other modes of dynamo operation that arise naturally in accretion disks -- at different magnetic Prandtl numbers, for example -- is a fascinating and completely open question.''

In the 25 years or so since then, there have been many shearing box simulations of the accretion disc dynamo, using a variety of codes, both grid-based and spectral, and a variety of physical parameters, with steadily increasing sophistication and resolution. The net result has been a much deeper understanding of the inner workings of the shearing box dynamo process, but the conclusions are unchanged. The simulations that assume zero externally imposed net magnetic flux typically find values of $\alpha$ at most around $\alpha \approx 0.01$ and often less. For example, \cite{Fromang:2007aa} using both grid-based and spectral codes obtained $\alpha \approx 0.001$ in their best resolved simulations, \cite{Heinemann:2009aa} use a spectral code and find $\alpha \approx 0.005$, \cite{Davis:2010aa} using a grid-based code find $\alpha \approx 0.01$, \cite{Salvesen:2016aa} using a grid-based code find $\alpha \approx 0.01$, \cite{Walker:2016aa} using a spectral code find that with no net imposed flux dynamo activity decays, so that $\alpha = 0$, and \cite{Mamatsashvili:2020aa} using a spectral code find $\alpha \approx 0.006 - 0.01$. It is clear that such low values are not in agreement with observations. However, there is a discussion of this tension in \cite{Shi:2016aa} who present grid-based shearing box simulations. There they find that higher values ($\alpha \approx 0.1$) can be obtained in an unstratified shearing box with an unusually large height-to-width ratio. Whether these larger values of alpha in tall, unstratified boxes carries over to the more realistic stratified case has been questioned \citep[e.g.][]{Ryan:2017aa}.

In summary, while such numerical simulations have successfully demonstrated the existence of a self-sustaining disc dynamo, they have not been able to produce magnetic fields of sufficient magnitude to agree with the observational data. Typically they produce values of $\alpha \approx 0.01$ or less, much smaller than the values of $\alpha \approx 0.2 - 0.3$ required to account for the observational data. So far, there is no explanation of why the value of $\alpha \approx 0.01$ emerges from the majority of the shearing box dynamo simulations. A physical understanding of what gives rise to the saturation of the dynamo process in numerical simulations would be of great value in understanding this difference between simulated and observed accretion discs.

\subsection{Global simulations}
\label{global}
\cite{King:2007aa} drew attention to the discrepancy between the magnitudes of the disc magnetic fields that are required by observations and those that can be achieved in numerical, shearing box  simulations. They discussed various reasons as to why the numerical simulations at that time might be inadequate. They noted that the shearing box approximation limits azimuthal structures to azimuthal wavenumbers $m=0$ or $m \gtrsim R/H \gg 1$, and suggested that this might play a role in suppressing large-scale azimuthal fields. For example, \cite{Tout:1992aa} note that in their semi-analytic dynamo model, which produces fields with $\beta \sim 1$, magnetic buoyancy plays an important role in the conversion of toroidal field to poloidal and acts at toroidal wavelengths $\lambda_\phi \gg H$. Thus it may be that it is the localisation of dynamo activity in the shearing box approximation that is responsible for the small values of $\alpha$ that are obtained in such simulations. We therefore need to consider global simulations.

We consider two recent global, grid-based, numerical simulations, presented by \cite{Pjanka:2020aa} and \cite{Oyang:2021aa}, which are specifically designed to simulate the accretion discs in cataclysmic variables. Our main interest here is the numerical diffusivity present in such simulations. In both of these simulations material is introduced at the outer edge of the disc (to model the stream of material from the low mass star) and is taken to contain small loops (size $ \sim H$) of magnetic flux, of magnitude $\beta \gg 1$ which can then initialise dynamo action. This seems reasonable, since the low mass star is sun-like and so presumably has surface magnetic fields \citep{Livio:1994aa}. \cite{Oyang:2021aa} also experiment with initially introducing small loops of flux at all radii in the disc. 

The disc thicknesses in dwarf novae are typically such that $H/R \approx 0.02 - 0.05$ \citep[see, for example,][]{Frank:2002aa}. The thinner disc in the simulations of \cite{Pjanka:2020aa} has $H/R = 0.1$ and for that disc they find in the body of the disc that $\alpha \approx 0.003$. The disc in \cite{Oyang:2021aa} has $H/R = 0.044$, and they find values of  $\alpha \approx 0.01$. Thus, neither of these global disc dynamo simulations is capable of accounting for the observed values of $\alpha$.

\section{What is missing?}
\label{missing}
We have seen that the shearing box approach does not seem able to produce a saturation level to dynamo action which involves strong enough magnetic fields to agree with the observational constraints. We have noted that one possibility for this might be that the shearing box approximation itself is too small scale (particularly in the azimuthal direction) to be able to provide an accurate description of dynamo activity in a global accretion disc.

However, we have also noted that those global disc simulations presented so far are also unable to produce the necessary saturation level to explain the observed dynamo action. We speculate here that in the global models this might come about for a different reason. The dominant physical process by which shear energy is converted into magnetic energy in an accretion disc is by the $R \phi$ shear acting on the radial $B_R$ field and converting it to azimuthal field, $B_\phi$. It is this process that ultimately drives the dynamo. If that physical process is artificially restricted, for example by too high a magnetic diffusivity, then the saturation level of dynamo activity might also be restricted. (We thank one of the referees for pointing out that this consideration may also be an issue for shearing box simulations.)

Numerical simulations are, of course, an approximation to reality. One way in which numerical simulations differ from reality is the inescapable inclusion of numerical dissipation, e.g. numerical viscosity and numerical magnetic diffusivity that are present in all numerical calculations (either explicitly as additional terms in the equations of motion or implicitly through e.g. truncation error at the grid scale). Because of these effects it is not possible to provide a solution to the equations of inviscid hydrodynamics with a numerical approach, as the calculation will necessarily involve some level of numerical viscosity. Similarly, it is not possible to solve the equations of ideal MHD with a numerical approach, as the calculation will necessarily involve numerical magnetic diffusivity. However, it seems to be typical to ``solve the equations of ideal MHD" with a numerical code without providing discussion on, or estimates of, the magnitude of these numerical effects \citep[see, for example,][]{Pjanka:2020aa,Oyang:2021aa}. (Note that, in contrast, e.g. \citealt{Fromang:2007aa,Walker:2016aa,Gogichaishvili:2017aa,Mamatsashvili:2020aa} use spectral codes in which they explicitly include viscosity and magnetic diffusivity of sufficient magnitude that their effects can be resolved numerically. However, such spectral codes are not readily applicable to global disc simulations. See also, e.g., \citealt{Fromang:2007aa} for the use of explicit dissipation in non-spectral approaches.) It is of course possible to make estimates of the numerical viscosity and magnetic diffusivity. The magnitudes of these depend typically on the sizes of the grid cells, the timesteps, and the orders of the numerical code in both space and time \citep[see, for example,][]{Rembiasz:2017aa}. But it is rare that sufficient information is presented for such estimates to be made by the reader.

As an illustrative example, we consider the evolution of a loop of magnetic flux of size $\approx H$ in the numerical codes employed in these global simulations. The simple question we ask is: what is the maximum possible flux amplification that can be achieved in these codes, ignoring any instabilities (such as MRI, buoyancy etc.) which might limit the effect. Ideally one would make use of the detailed code quantities, together with formulae for numerical magnetic diffusivity, $\eta_N$, such as those proposed by \cite{Rembiasz:2017aa}. However, not enough information is provided in these papers to undertake such an analysis. For this reason we shall content ourselves with the following approximation.

For the loop we are considering, the distance in the azimuthal direction over which the radial component of $\bf B$ reverses is $\approx H$. After a time $t$, the angle an initially radial field makes with the azimuthal direction is $\approx (U^\prime t)^{-1}$. If the radial grid cell has size $\Delta R$, then after a time $t_{\rm max}$, where $U^\prime t_{\rm max} = H/(2 \Delta R)$, the grid cell should in principle contain magnetic fluxes of opposing signs. We may assume that at this point numerical magnetic diffusivity ensures that this does not happen. We also note that at that time, the initial magnetic field has been amplified by a factor of $\approx U^\prime t_{\rm max}$. Thus, as a first approximation we may assume that in an Eulerian numerical code, the maximum amplification of the field energy is given approximately by
\begin{equation}
{\mathcal E}_B(t_{\rm max})  / {\mathcal E}_B(0) = (B(t_{\rm max}) / B_0)^2 \approx \left(\frac{H}{2 \Delta R}\right)^2\,.
\end{equation}

The codes used by \cite{Pjanka:2020aa} and \cite{Oyang:2021aa} employ, respectively, adaptive mesh refinement (AMR) and static mesh refinement (SMR), and thus have variable cell size. In \cite{Pjanka:2020aa} for the thinner disc case (which has disc thickness closer to that required for modelling dwarf nova discs),  they take $H = 0.1 R$, The initial cells have $\Delta R/R = 0.1$, and there are up to 5 levels of cell refinement. Thus, in principle,  $\Delta R$ might be reduced by a factor of up to 32. In their model this would imply $\Delta R/R \approx  0.003$. From this we deduce that the maximum achievable enhancement of magnetic field energy is $\approx 16^2$. In \cite{Oyang:2021aa} there are up to 256 cells logarithmically spaced between the inner radius $r=1$ and the outer radius $r=23.4$. This gives a smallest radial cell size of $\Delta R/R \approx 0.012$, to be compared with the disc scaleheight $H/R = 0.044$. Thus here the maximum field energy enhancement is by a factor of $\approx 4$.

In order to relate these numbers to the actual numerical viscosities present in the codes we now present a formal computation of the evolution of such field loops in a simple linear shear flow.

\section{Formal computation of loop evolution}
\label{example}
To illustrate the effect of magnetic diffusivity on magnetic field amplification, we consider a simple two-dimensional flow with differing initial magnetic field configurations. We have noted that the fundamental mechanism by which magnetic energy is increased, by tapping the energy available in the background  $R\phi-$shear flow, is the conversion (stretching) of radial magnetic field to form azimuthal field. In \cite{Pjanka:2020aa} and \cite{Oyang:2021aa} small loops of magnetic flux are used to seed the radial field. Alternatively one may think of the effect of the MRI on an initially azimuthal field, as producing radial undulations of the field, which can be thought of as loops of flux in the $R\phi-$plane. Such loops would then be stretched by the underlying $R\phi-$shear flow.

To contrast with the usual shearing box nomenclature (in which rotation, i.e. coriolis forces, are included), we shall consider two dimensions, but take the $x$-coordinate to correspond to the ``azimuthal'' direction and the $y$-coordinate to the ``radial'' direction. Thus, in the $xy$-plane we assume an inexorable linear shear flow of the form ${\bf u} = (U^\prime y, 0)$. with $U^\prime$ a constant. First we consider the simple case where the initial field has only a component in the $y$-direction. We then consider the evolution of initial magnetic field loops. The exact general solution for arbitrary initial conditions can be found in Appendix~\ref{appA}.

\subsection{Straight (radial) field lines}
We take an initial magnetic field, at time $t=0$ of the form ${\bf B} = B_0 (0, \cos (kx) )$. This can be thought of as an approximation to a magnetic loop of size $l = \pi/k$ in a shearing (but not rotating) box of size $\sim l$. We assume the fluid to be incompressible, and to obey the standard MHD equations with a magnetic diffusivity $\eta$.

In this case we can define the magnetic flux function $A(x,y, t)$ such that \citep{Davidson:2001aa,Yeates:2011aa}
\begin{equation}
{\bf B} = \left( \frac{\partial A}{\partial y}, - \frac{\partial A}{\partial x} \right)\,,
\end{equation} 
which in this case obeys the equation
\begin{equation}
\label{Aevolve}
\frac{\partial A}{ \partial t} + U^\prime y \frac{\partial A}{\partial x} = \eta \nabla^2 A\,.
\end{equation}
The flux function is such that the magnetic field lines are given by $A = $ const.

At time $t=0$ we have that 
\begin{equation}
A(x,y,t=0) =  - k^{-1} B_0 \sin (kx)\,.
\end{equation}

If we have ideal MHD, so that $\eta = 0$, we expect the field lines to be simply advected by the shear flow, so that
\begin{equation}
A(x,y,t) =  - k^{-1} B_0 \sin k(x - U^\prime t y)\,. \label{eta0}
\end{equation}
In this case, the magnetic field grows indefinitely in a linear fashion, viz.,
\begin{equation}
{\bf B} = B_0 (U^\prime t, 1) \cos k (x - U^\prime ty)\,.
\end{equation}

For non-ideal MHD, with magnetic diffusivity  $\eta > 0$, we can see that the solution is separable, and of the form
\begin{equation}
A = f(t) \sin k(x - U^\prime ty)\,.
\end{equation}

Substituting this into Equation~\ref{Aevolve} we find that
\begin{equation}
\frac{df}{dt} = - \eta f k^2 [1 + (U^\prime)^2 t^2]\,.
\end{equation}

Thus we have that 
\begin{equation}
A(x,y,t) = - k^{-1} B_0 \sin [k(x - U^\prime ty)] \exp \left\{- \eta k^2 [t + \frac{1}{3} (U^\prime)^2 t^3] \right\}\,,
\end{equation}
and, therefore, that
\begin{equation}
{\bf B} = B_0 (U^\prime t, 1) \cos [k(x-U^\prime ty)]  \exp \left\{- {\mathcal R}_{\rm m}^{-1}  [U^\prime t + \frac{1}{3} (U^\prime t)^3] \right\}\,,
\end{equation}
where ${\mathcal R}_{\rm m} = U^\prime/(k^2 \eta)$ is the relevant magnetic Reynolds number.

Thus the spatially averaged (e.g. over a box of size $-l \le x,y \le l$) value of the magnetic energy (${\mathcal E}_B$) is of the form \citep[cf.][]{Pringle:2017aa}
\begin{equation}
{\mathcal E}_B(t)  / {\mathcal E}_B(0) = [1 + (U^\prime t)^2] \exp \left\{- 2{\mathcal R}_{\rm m}^{-1}  [U^\prime t + \frac{1}{3} (U^\prime t)^3] \right\}\,. \label{enerlines}
\end{equation}

We plot this in Figure~\ref{fig1}, for various values of ${\mathcal R}_{\rm m}$. From this we can see that the evolution is as follows: 
\begin{figure}
  \centerline{\includegraphics[width=0.4667\textwidth]{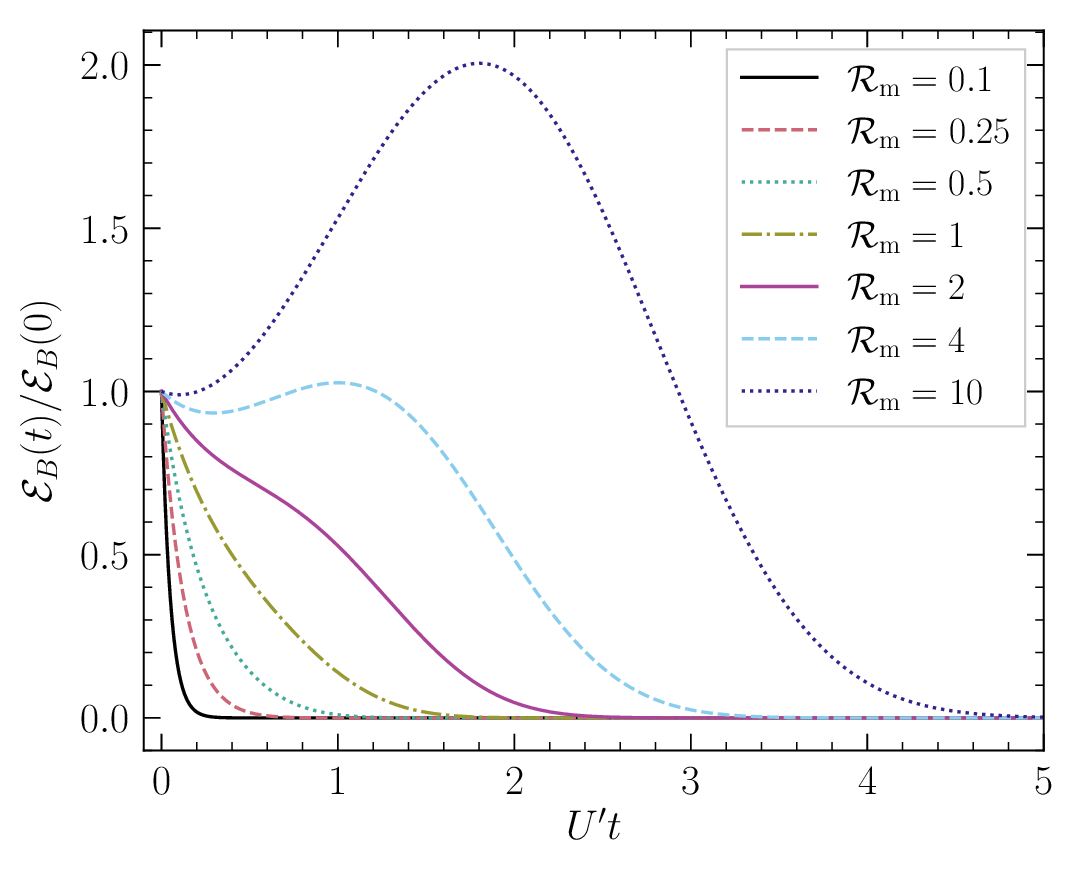}\hfill\includegraphics[width=0.499\textwidth]{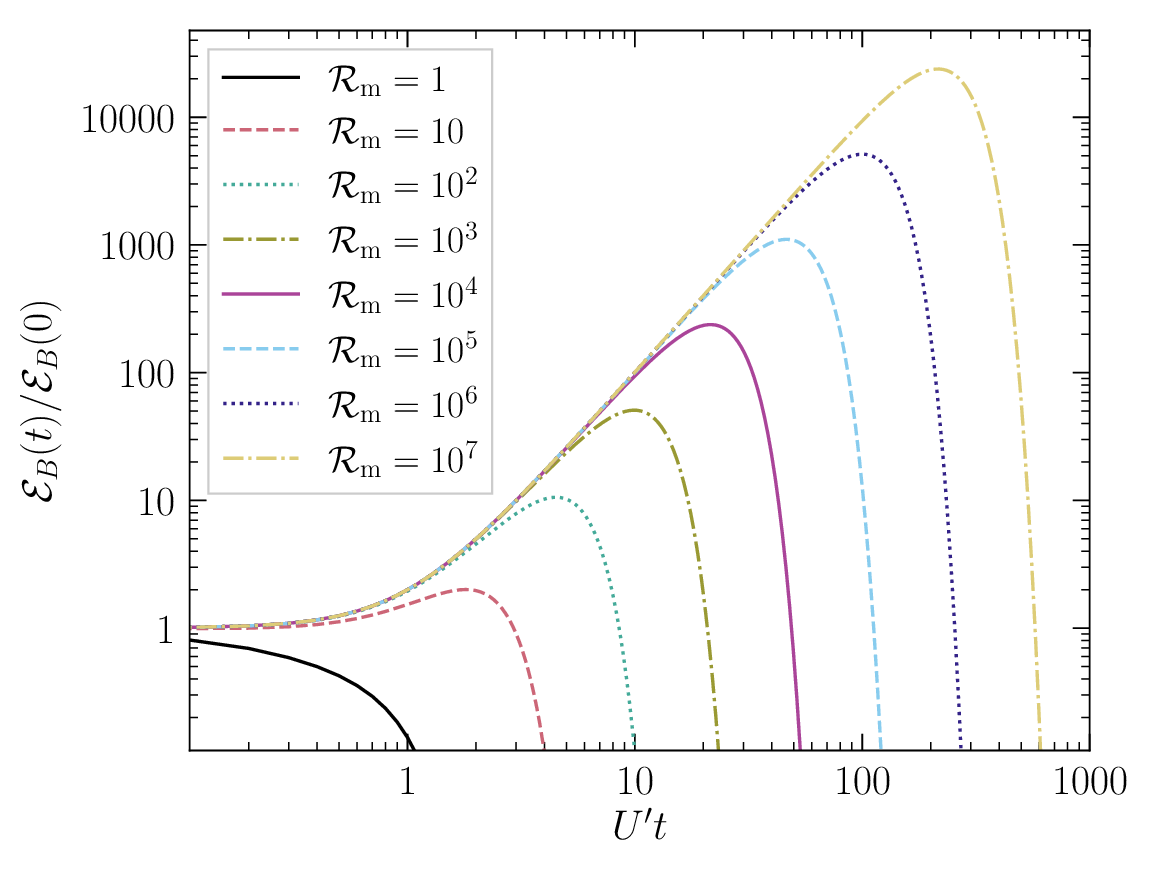}}
  \caption{Evolution of the magnetic energy, scaled to the initial value, with time for several values of the magnetic Reynolds number, ${\mathcal R}_{\rm m}$, for the case of initial radial field lines (equation~\ref{enerlines}). Left panel shows values of $0.1 \le {\mathcal R}_{\rm m} \le 10$ with the magnetic energy on a linear scale. Right panel shows values of ${\mathcal R}_{\rm m}$ up to $10^7$ with the magnetic energy on a log scale. For small ${\mathcal R}_{\rm m}$ the energy decays rapidly, while for large ${\mathcal R}_{\rm m}$ the field initially decays before exhibiting growth and final decay.}
\label{fig1}
\end{figure}

\begin{enumerate} 
\item Initially the magnetic field energy starts to decay at a rate of $2U^\prime/{\mathcal R}_{\rm m}$ due to magnetic diffusivity. This initial decay phase is brief.
\item Provided that ${\mathcal R}_{\rm m}$ is large enough, the shear is strong enough to provide an enhancement of the magnetic field strength. For large ${\mathcal R}_{\rm m}$, field growth occurs after a time $U^\prime t \approx {\mathcal R}_{\rm m}^{-1}$. In this phase the magnetic energy grows linearly and this is a direct consequence of the shear.
\item Eventually, the shear decreases the spatial length-scale of the magnetic field sufficiently that diffusivity wins, and, as is well known, the field ultimately decays \citep[cf.][]{Weiss:1966aa,Moffatt:1978aa}. The exponential decay is quicker than the simple $\exp (- \eta k^2 t)$, which is what happens if there is no shear, because the combination of shear and diffusivity hastens the process.
\end{enumerate}

\begin{figure}
  \centerline{\includegraphics[width=0.5\textwidth]{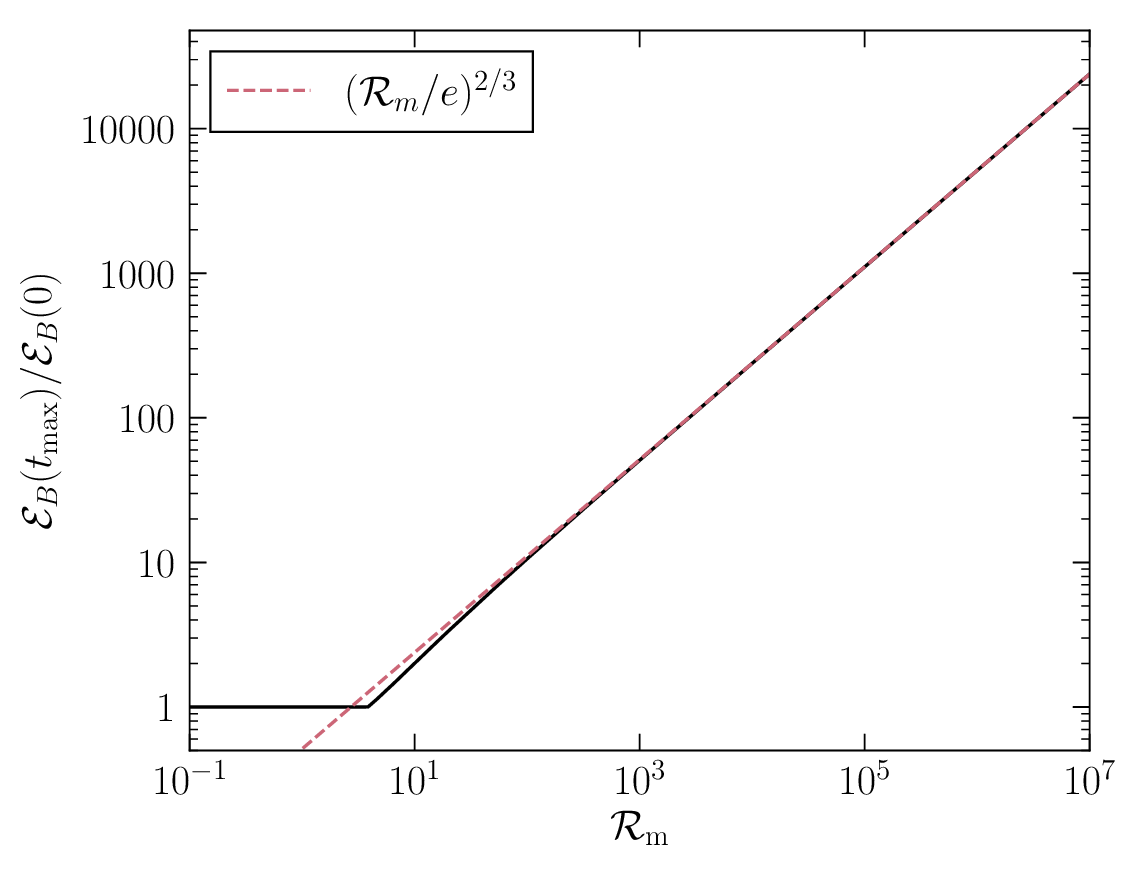}}
  \caption{The maximum growth factor of the magnetic field energy plotted as a function of the magnetic Reynolds number, ${\mathcal R}_{\rm m}$, for the case of initial radial field lines (i.e. the maximum value attained from equation~\ref{enerlines}). The red-dashed line indicates the prediction appropriate to the limit of large ${\mathcal R}_{\rm m}$, which is accurate for ${\mathcal R}_{\rm m} \gtrsim 10$. For ${\mathcal R}_{\rm m} \lesssim 3.8$ the magnetic energy never grows back above the original value.}
\label{fig2}
\end{figure}

The maximum possible enhancement of the magnetic field energy, ${\mathcal E}_B(t_{\rm max})/{\mathcal E}_B(0) $, depends on ${\mathcal R}_{\rm m}$. We plot this in Figure~\ref{fig2}. For ${\mathcal R}_{\rm m} \gg 1$, the maximum enhancement occurs when $U^\prime t \approx {\mathcal R}_{\rm m}^{1/3}$, and is equal to ${\mathcal E}_B(t_{\rm max})/{\mathcal E}_B(0) \approx ({\mathcal R}_{\rm m}/e)^{2/3}$.

\begin{figure}
  \centerline{\includegraphics[width=0.475\textwidth]{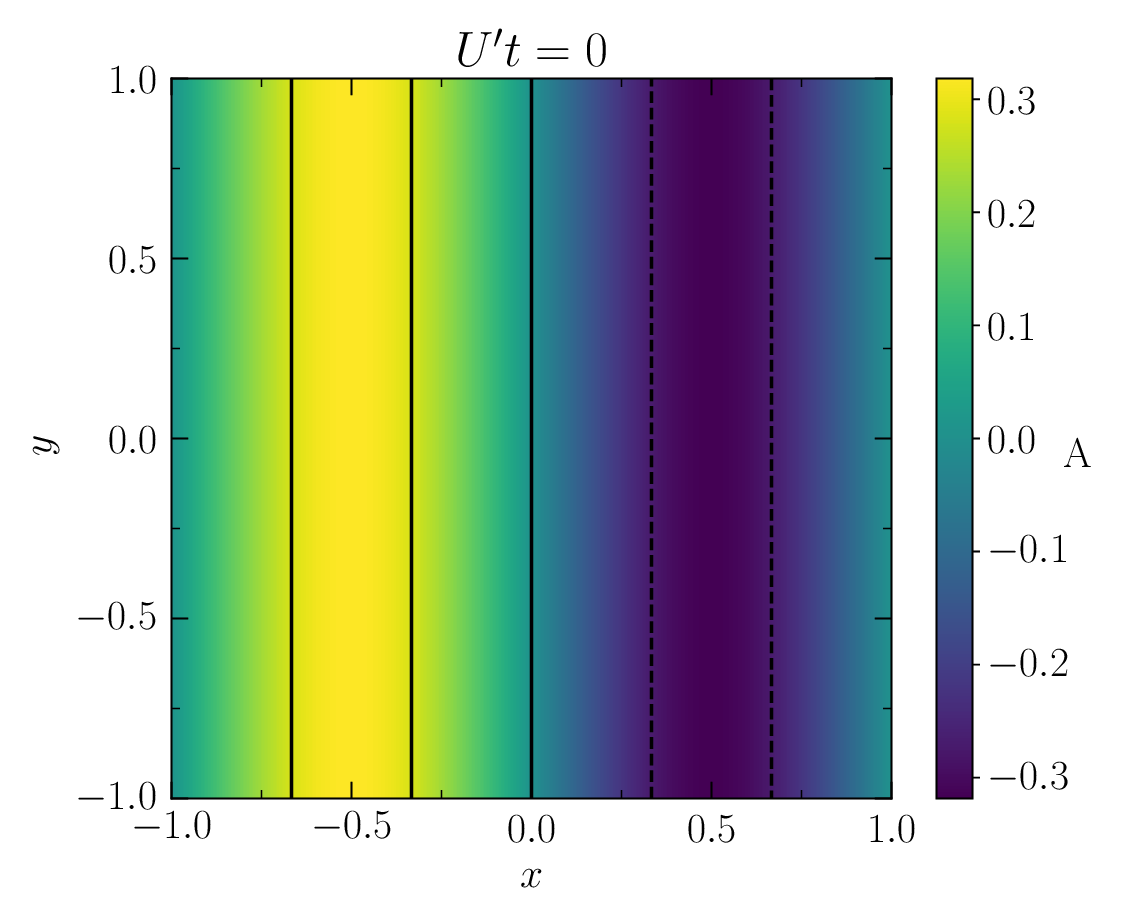}\hfill\includegraphics[width=0.475\textwidth]{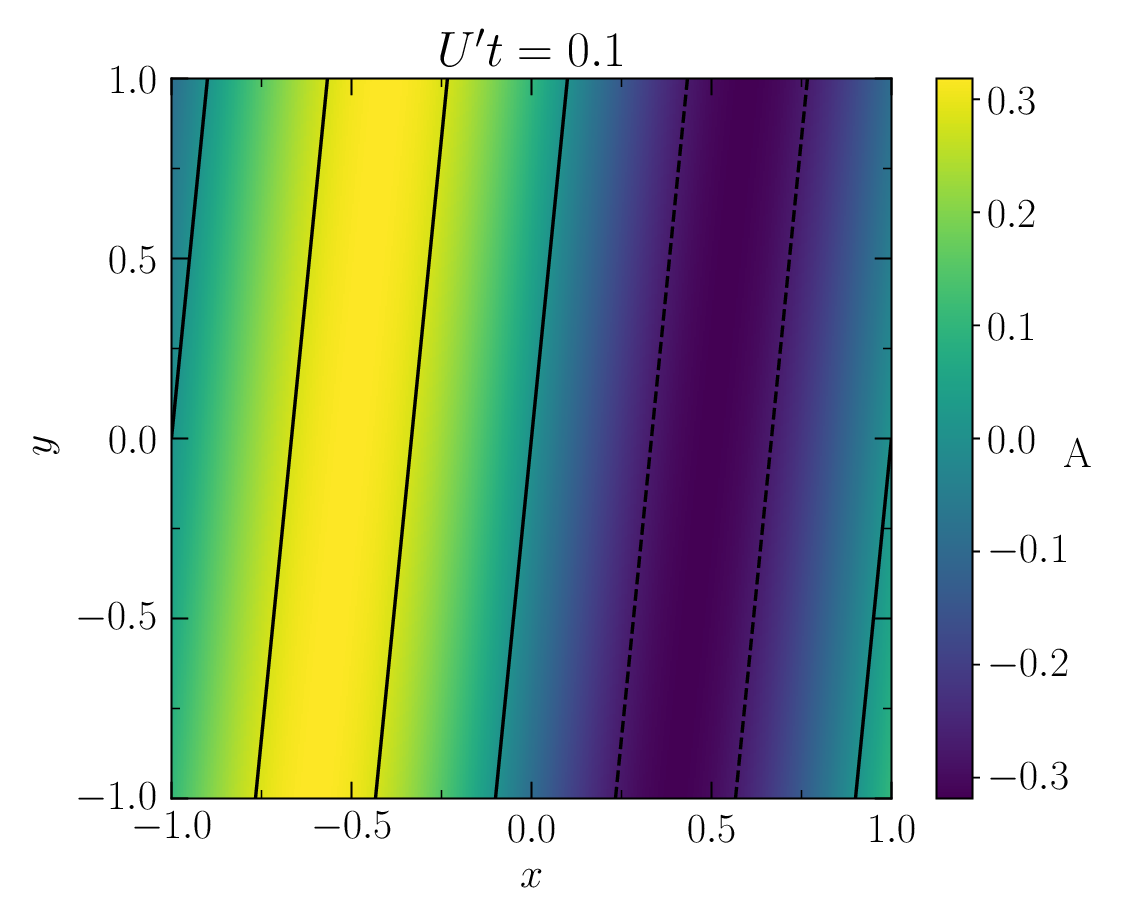}}
  \centerline{\includegraphics[width=0.475\textwidth]{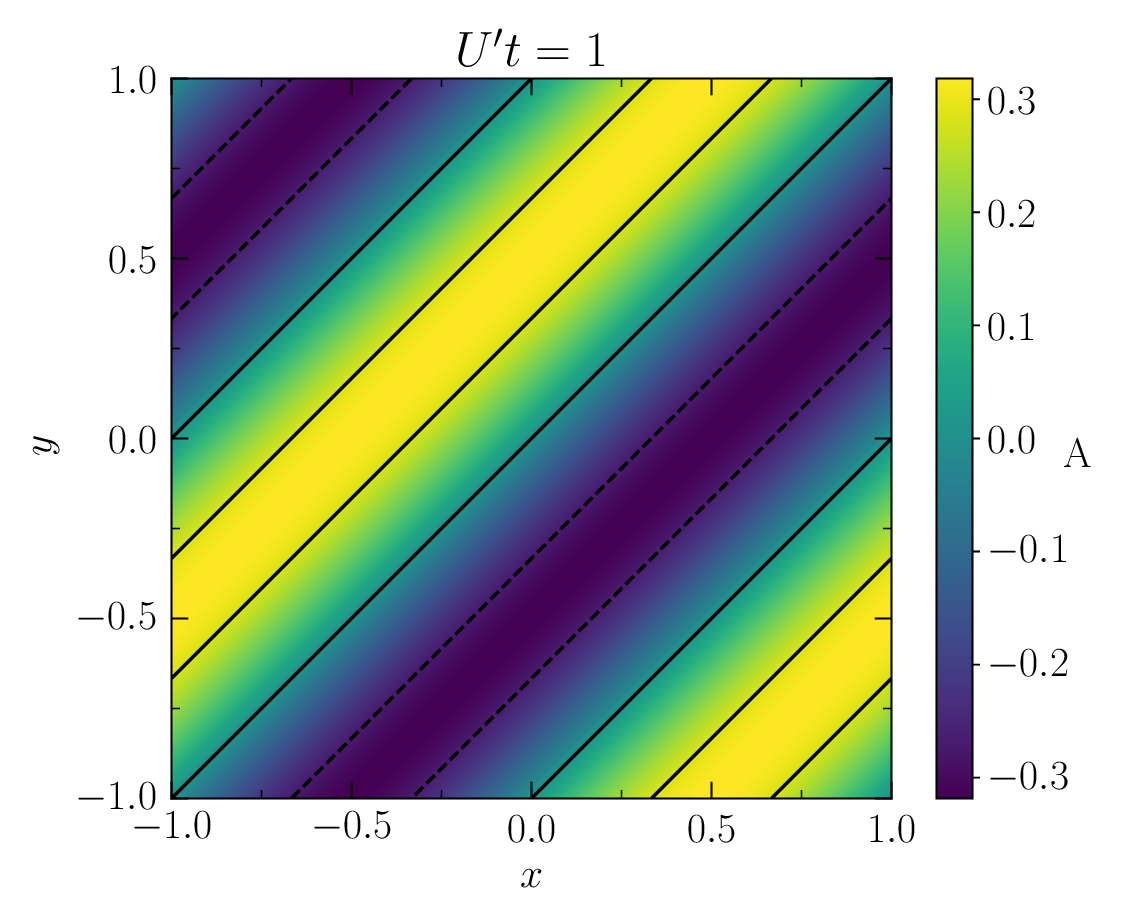}\hfill\includegraphics[width=0.475\textwidth]{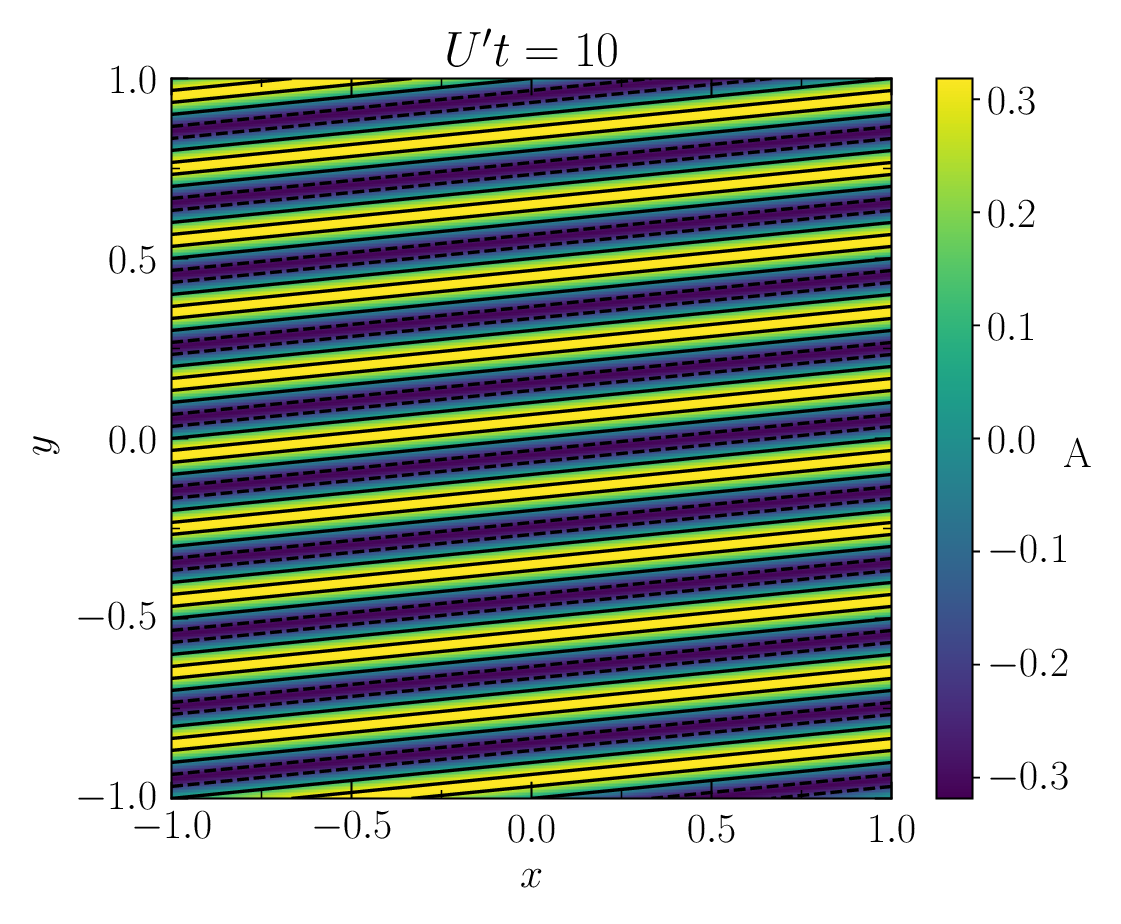}}
  \caption{Evolution of the magnetic field lines from an initially linear radial ($y$-direction) configuration. From left to right, and then top to bottom, time increases from $U^\prime t = 0$ (upper left panel) up to $U^\prime t = 10$ (lower right panel). The colour bar denotes the value of $A$, with lines of constant $A$ delineating the field lines. For clarity, we also plot 5 field lines, which at $U^\prime t = 0$ correspond to $A = -k^{-1}B_0 \sin (k\gamma)$ with $k=\pi$, $B_0=1$, and $\gamma = -2/3, -1/3, 0, 1/3, 2/3$ (the solid black lines represent those field lines with $A \ge 0$ and the dashed lines represent those with $A < 0$). In subsequent panels only these field lines are plotted, with the apparent number increased due to stretching of the field lines in the $x$-direction. This figure illustrates the conversion of radial field ($y$-direction) into azimuthal field ($x$-direction). The strength of the field varies in time and the evolution of the strength depends on ${\mathcal R}_{\rm m}$ (see Figure~\ref{fig1}). Here we have plotted the case where where $\eta \rightarrow 0$ (corresponding to ${\mathcal R}_{\rm m} \rightarrow \infty$), and as such the range of values for $A$ remains fixed (cf. equation~\ref{eta0}).}
\label{fig3}
\end{figure}

In Figure~\ref{fig3} we plot the geometric evolution of the magnetic field lines at different times as an illustration; note that the value of ${\mathcal R}_{\rm m}$ does not affect the geometry and only affects the field strength in this case. From left to right the panels correspond to times of $U^\prime t = 0,0.1,1,10$.

\subsection{Loops of magnetic field}
We now explore the behaviour of loops of magnetic flux. To stay with one Fourier mode in each direction, we take
\begin{equation}
A(x,y,t=0) = \sin k_x x \sin k_y y\,.
\end{equation}

In the region $0 \le k_x x,\, k_y y \le \pi$  the magnetic field takes the form of loops around the point $(\pi/2, \pi/2)$ with the field pointing in the anti-clockwise direction. The rest of space is populated with similar rectangular tiles containing loops of magnetic field of alternating chirality. 

We note that we may rewrite the flux function as
\begin{equation}
A(x,y,t=0) = \frac{1}{2} [ \cos (k_x x- k_y y) - \cos (k_x x+k_y y) ]\,.
\end{equation}

Thus with zero diffusivity we have simple advection of the field by the shear so that
\begin{equation}
A(x,y,t) = \frac{1}{2} [ \cos (k_x x - U^\prime t k_x y - k_y y) - \cos (k_x x - U^\prime t k_x y + k_y y)]\,.
\end{equation}

Because we have have a combination of two Fourier modes, for $\eta > 0$ we now look for solutions of the form
\begin{equation}
A(x,y,t) = \frac{1}{2} [ f_1(t) \cos (k_x x - U^\prime t k_x y - k_y y) - f_2(t) \cos (k_x x - U^\prime t k_x y + k_y y)]\,,
\end{equation}
which, as before, can be substituted into the evolution equation (\ref{Aevolve}) and solve for $f_1(t)$ and $f_2(t)$. Using the initial conditions that at $t=0$, $f_1(0) = f_2(0) = 1$, we have
\begin{equation}
f_1(t) = \exp \left\{ - \eta \left[ (k_x^2 + k_y^2)t + k_x k_y U't^2 + \frac{1}{3}k_x^2 (U')^2 t^3 \right] \right\}\,,
\end{equation}
and
\begin{equation}
f_2(t) = \exp \left\{ - \eta [ (k_x^2 + k_y^2)t - k_x k_y U't^2 + \frac{1}{3}k_x^2 (U')^2 t^3 ] \right\}\,,
\end{equation}
Putting this all together we have
\begin{equation}
\begin{split}
A(x,y,t) =\left[  \sin k_x(x - U'ty) \sin k_yy \cosh ( \eta k_x k_y U't^2)  
 -  \cos k_x (x - U'ty) \cos k_y y \sinh ( \eta k_x k_y U't^2) \right]  \\  
 \times \exp \left\{ - \eta [(k_x^2 + k_y^2)t + \frac{1}{3} k_x^2 (U')^2 t^3] \right \}\,.
 \end{split}
\end{equation}

In this case the evolution of the spatially averaged magnetic field energy (e.g. over a box of size $-\pi \le k_x x, k_y y \le \pi$) is
\begin{equation}
\frac{{\mathcal E}_B(t)}{{\mathcal E}_B(0)} = \frac{[k_x^2 + (k_y + k_x U't)^2] f^2_1(t) + [ k_x^2 + (k_y - k_x U't)^2] f^2_2(t)}{2 [k_x^2 + k_y^2]}\,. \label{loopener}
\end{equation}

\begin{figure}
  \centerline{\includegraphics[width=0.4667\textwidth]{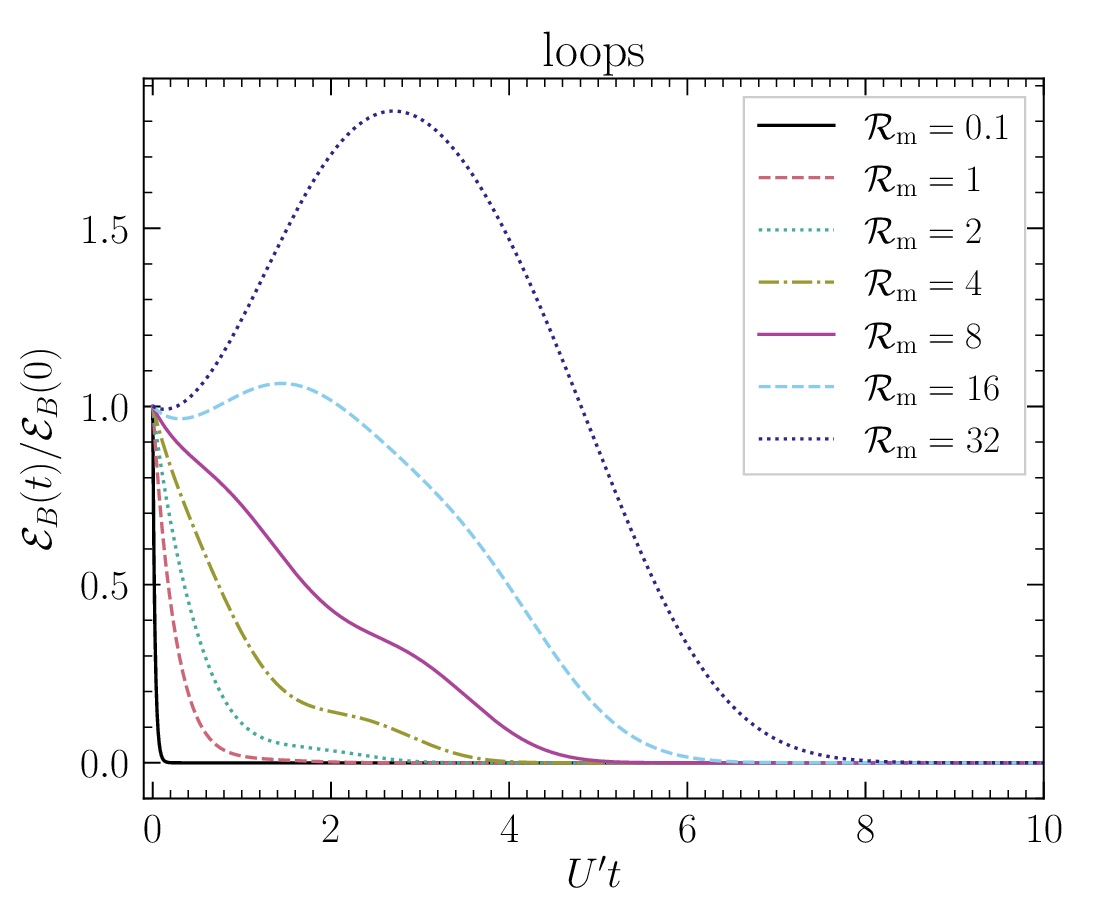}\hfill\includegraphics[width=0.499\textwidth]{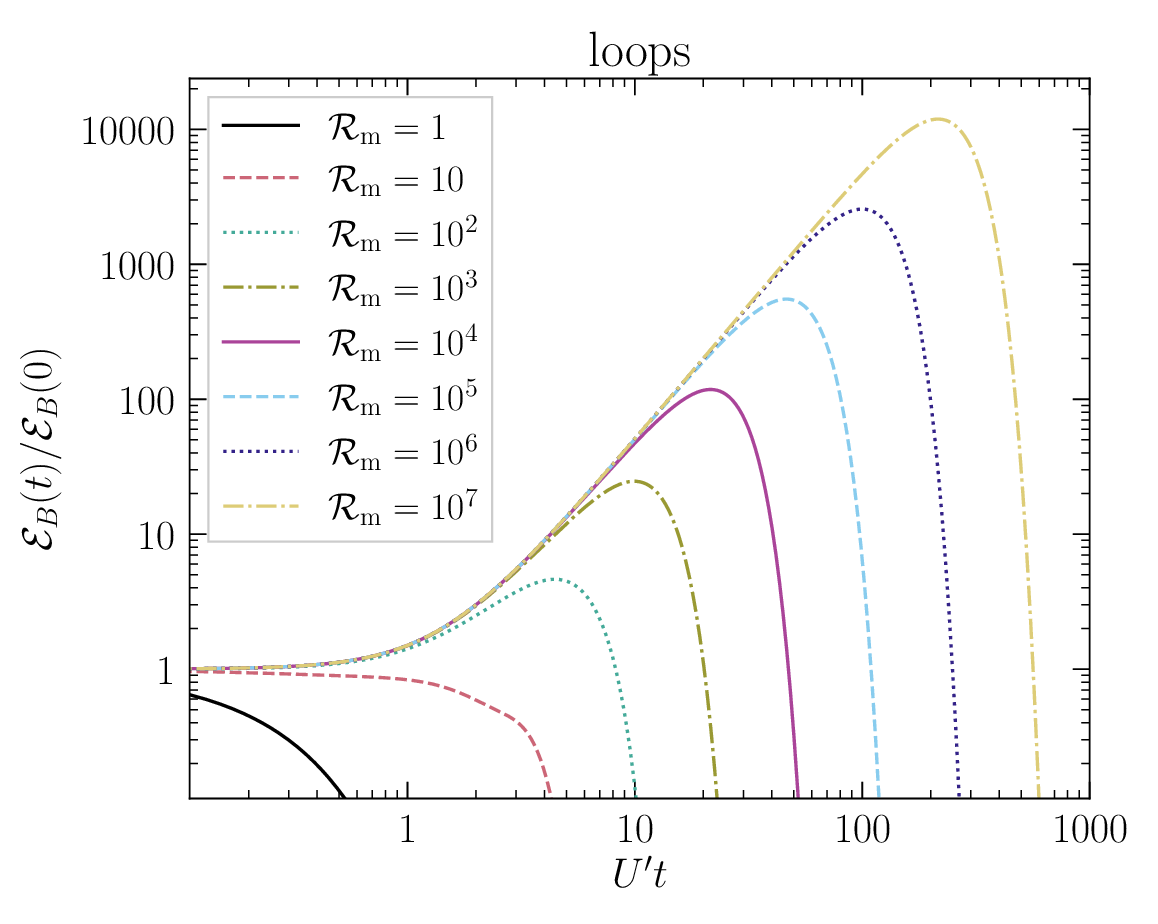}}
  \caption{Evolution of the magnetic energy, scaled to the initial value, with time for several values of the magnetic Reynolds number, ${\mathcal R}_{\rm m}$, for the case of initial loops of magnetic field (equation~\ref{loopener}). Left panel shows values of $0.1 \le {\mathcal R}_{\rm m} \le 32$ with the magnetic energy on a linear scale. Right panel shows values of ${\mathcal R}_{\rm m}$ up to $10^7$ with the magnetic energy on a log scale. For small ${\mathcal R}_{\rm m}$ the energy decays rapidly, while for large ${\mathcal R}_{\rm m}$ the field initially decays before exhibiting growth and final decay. The behaviour is similar, particularly at large ${\mathcal R}_{\rm m} \gg 1$ with larger ${\mathcal R}_{\rm m} \gtrsim 14.7$ required to exhibit field energy growth.}
\label{fig4}
\end{figure}

As before, we define the magnetic Reynolds number as ${\mathcal R}_{\rm m} = U^\prime/(k_x^2\eta)$, but here a secondary number ($k_y/k_x$) is required to define the flow. For illustration we make the following figures with $k_y/k_x = 1$. First we plot the evolution of the magnetic energy (equation~\ref{loopener}) in Figure~\ref{fig4}. The result is similar to that for initial straight field. For small ${\mathcal R}_{\rm m}$ the field energy decays rapidly. For large ${\mathcal R}_{\rm m}$ there is a period of significant growth of the field energy before the field later decays. For loops the critical ${\mathcal R}_{\rm m}$ at which the field exhibits any growth is significantly larger than for the straight field lines, requiring ${\mathcal R}_{\rm m} \gtrsim 14.7$ for loops rather than just ${\mathcal R}_{\rm m} \gtrsim 3.8$ for lines.

\begin{figure}
  \centerline{\includegraphics[width=0.5\textwidth]{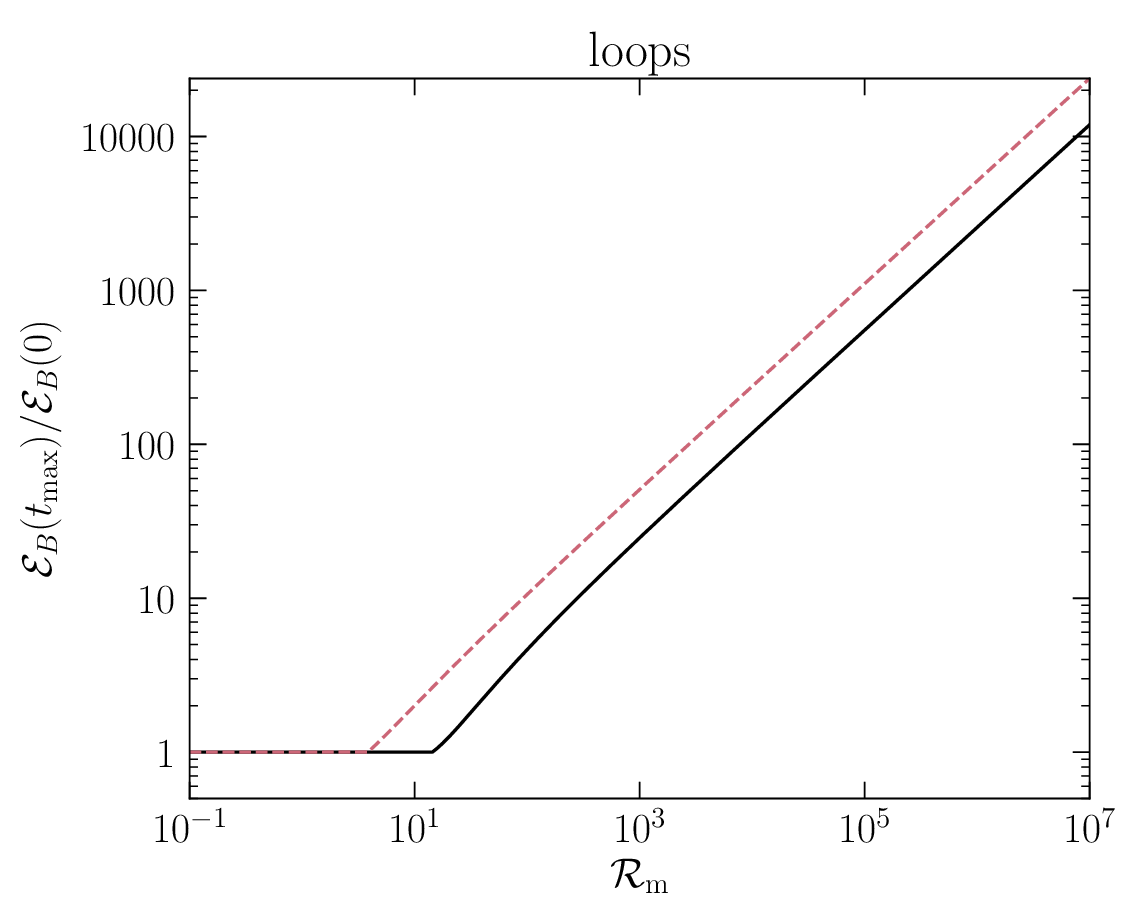}}
  \caption{The maximum growth factor of the magnetic field energy plotted as a function of the magnetic Reynolds number, ${\mathcal R}_{\rm m}$, for the case of initial loops of magnetic field (i.e. the maximum value attained from equation~\ref{loopener}). The red-dashed line shows the solution presented in Figure~\ref{fig2} for the case of initial radial field lines. Here, for ${\mathcal R}_{\rm m} \lesssim 14.7$ the magnetic energy never grows back above the original value, which contrasts with the value of ${\mathcal R}_{\rm m} \lesssim 3.8$ in the initial radial field line case.}
\label{fig5}
\end{figure}

\begin{figure}
  \centerline{\includegraphics[width=0.475\textwidth]{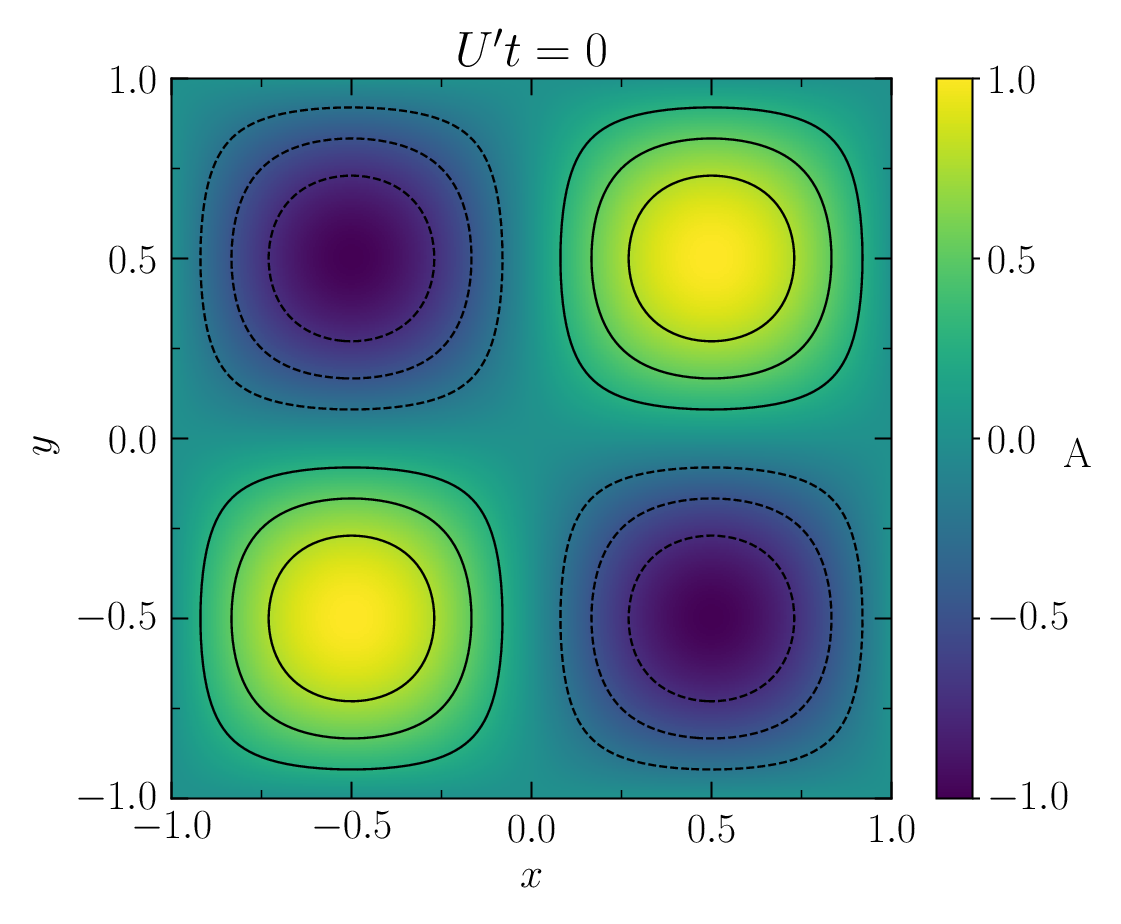}\hfill\includegraphics[width=0.475\textwidth]{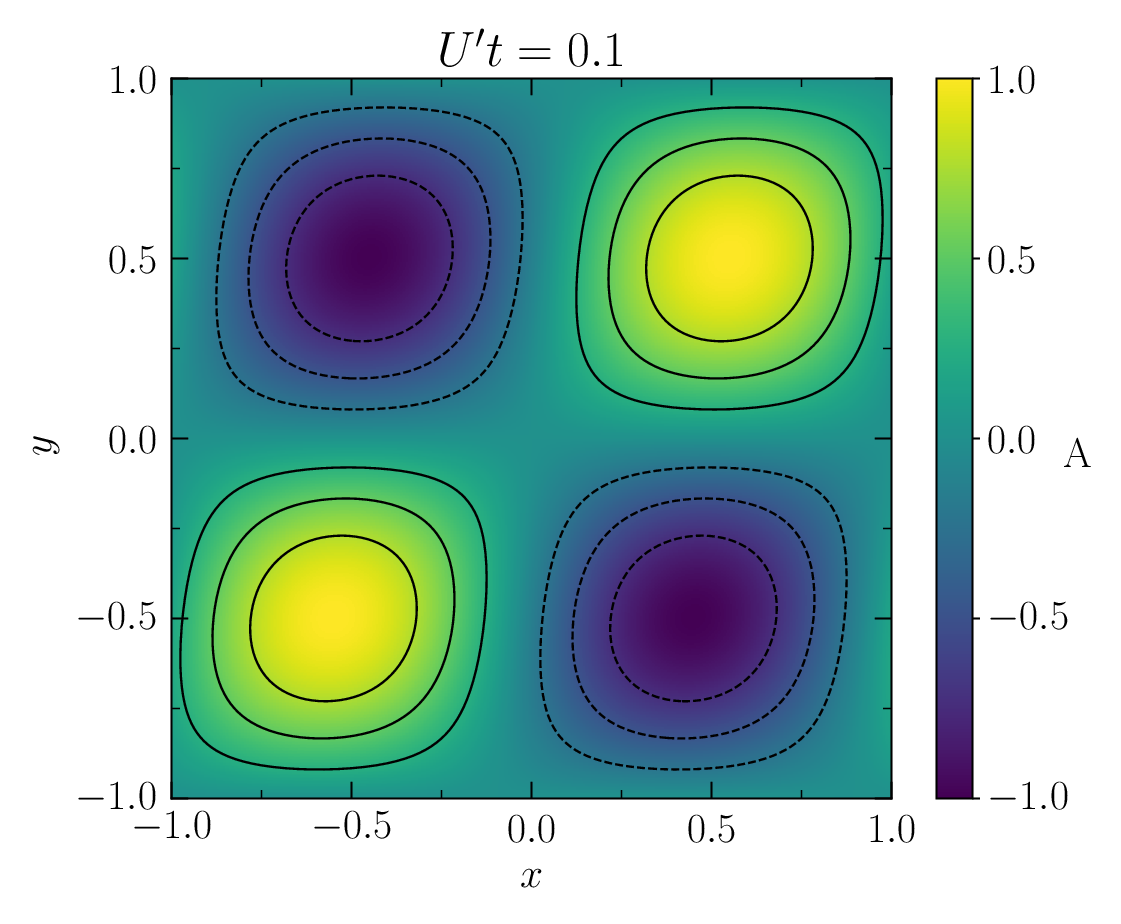}}
  \centerline{\includegraphics[width=0.475\textwidth]{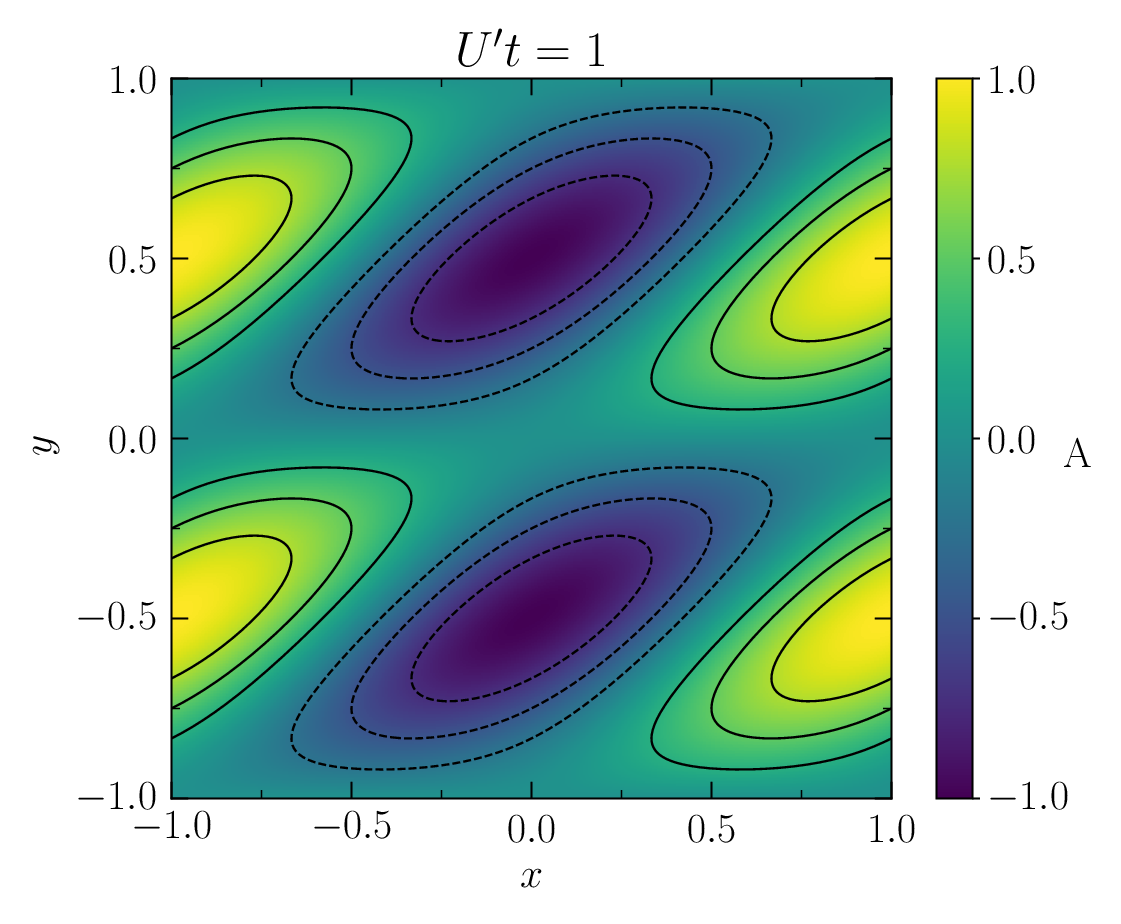}\hfill\includegraphics[width=0.475\textwidth]{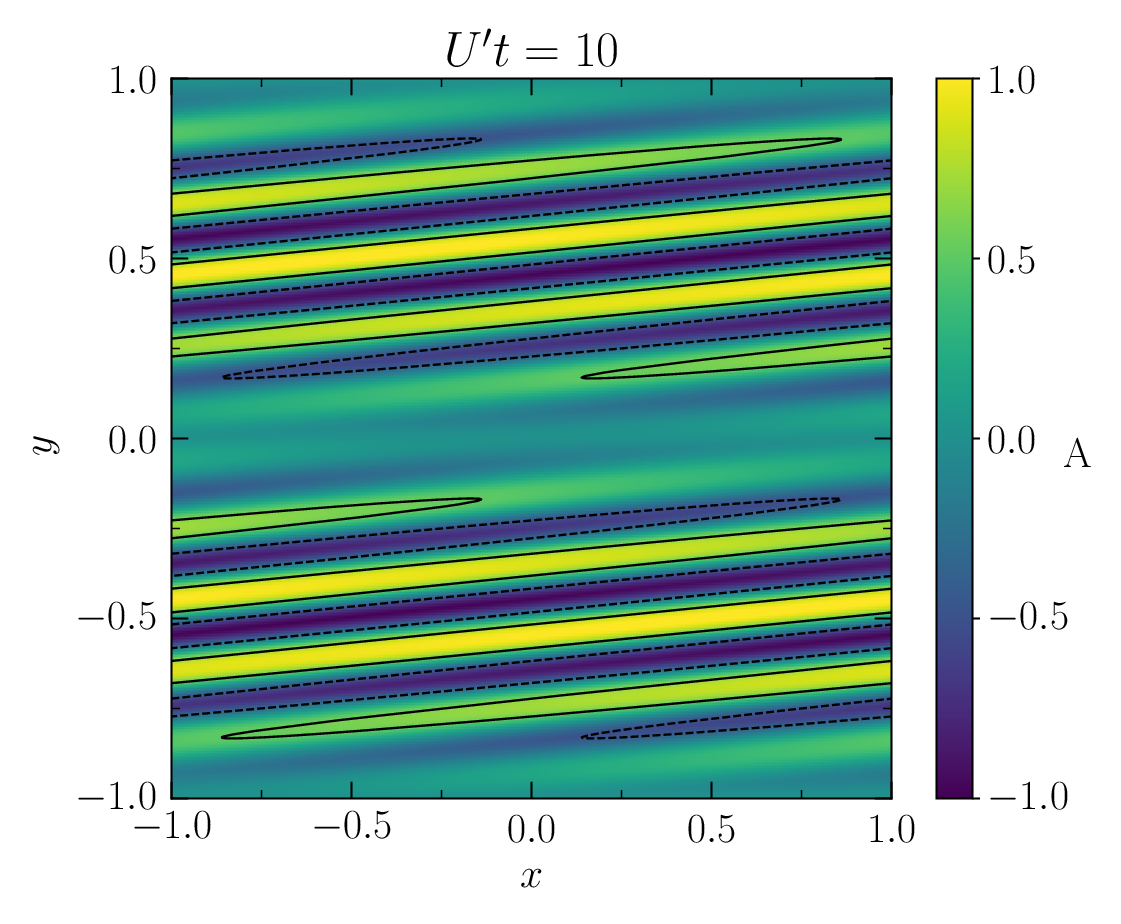}}
  \caption{Evolution of the magnetic field lines from initial loops of magnetic field with $k_x = k_y = \pi$. From left to right, and then top to bottom, time increases from $U^\prime t = 0$ (upper left panel) up to $U^\prime t = 10$ (lower right panel). The colour bar denotes the value of $A$, with lines of constant $A$ delineating the field lines. The regions with $A > 0$ (yellow) correspond to regions where the magnetic field lines are oriented in the counterclockwise direction, while for regions with $A < 0$ (blue) the magnetic field lines are oriented with the opposite chirality. For clarity, we also plot field lines corresponding to $A = -3/4, -1/2, -1/4,\,1/4,\,1/2,\,3/4$ (the solid black lines represent those field lines with $A \ge 0$ and the dashed lines represent those with $A < 0$) with the exception of the final panel where only field lines with $A=-1/2$ and $1/2$ are plotted for clarity. The strength of the field varies in time and the evolution of the strength depends on ${\mathcal R}_{\rm m}$ (see Figure~\ref{fig4}). Here we have plotted the case where where $\eta \rightarrow 0$ (corresponding to ${\mathcal R}_{\rm m} \rightarrow \infty$), and as such the range of values for $A$ remains fixed (cf. equation~\ref{eta0}). As with the initial linear field case (see Figure~\ref{fig3}), the field lines are stretched due to the shear. As time proceeds the solution for initial loops of field takes on a similar geometry to the case with initial lines of field for $U^\prime t \gg 1$.}
\label{fig6}
\end{figure}

In Figure~\ref{fig5} we plot the maximum growth of the magnetic energy against ${\mathcal R}_{\rm m}$. In this figure the black solid line corresponds to maximum growth from initial field loops, while the red-dashed lines is for the case of field lines plotted in Figure~\ref{fig2}. Here we can see that the maximum growth is weaker for loops of field compared to field lines for the same value of ${\mathcal R}_{\rm m}$. However, we can also see that both cases exhibit the same functional dependence of the maximum growth rate on ${\mathcal R}_{\rm m}$; for field loops with $k_x = k_y = 1$ the maximum growth in magnetic energy is ${\mathcal E}_B(t_{\rm max})/{\mathcal E}_B(0) \approx (1/2)({\mathcal R}_{\rm m}/e)^{2/3}$. This is to be expected as at late times, where the maximum is reached for large ${\mathcal R}_{\rm m}$, the initial field loops have already been sheared out into lines of field. To illustrate this we plot the magnetic field lines resulting from the initial loops of field in Figure~\ref{fig6}. This Figure again shows the conversion of radial to azimuthal field, and that for $U^\prime t \gg 1$ the field structure is similar to the case with initial lines of field as the loops become strongly stretched in the $x$-direction.

\section{Discussion \& Conclusions}
\label{discussion}
It has long been known that shearing box numerical simulations of the dynamo dynamics to be found in accretion discs do not provide an explanation of the observational data. In particular, for the highly ionized discs in outbursting dwarf novae, the dimensionless viscosity parameter exceeds that obtained in numerical simulations by more than an order of magnitude \citep{King:2007aa,Martin:2019aa}.  We note that in the spectral code implementation of the shearing box the magnetic diffusivity is controlled and magnetic Reynolds' numbers in the range $10^4 - 10^5$  can be achieved \citep{Fromang:2007aa,Walker:2016aa,Mamatsashvili:2020aa}. As we have seen, the physical value of the magnetic Reynolds' number in the observed dwarf nova discs is around $\sim 10^{10}$.  Thus, while the lack of agreement between the shearing box simulations and the observational data may yet be due to a discrepancy in magnetic diffusivity, it may also be that the reason is due to the shearing box approximation itself. If so, then the next step is to undertake global disc simulations. 

Such simulations have been recently presented by \cite{Pjanka:2020aa} and \cite{Oyang:2021aa}. However, of necessity, the numerical methods used by \cite{Pjanka:2020aa} and \cite{Oyang:2021aa} have intrinsic numerical magnetic diffusivities. We estimate (Section~\ref{missing}) that the maximum factor by which the field strength of a small magnetic loop can be enhanced in these codes is $\sim 16$, which corresponds to a factor of $16^2$ in the magnetic field energy. From Section~\ref{example} we see that ${\mathcal E}_B(t_{\rm max})/{\mathcal E}_B(0) \approx ({\mathcal R}_{\rm m}/e)^{2/3}$, suggesting that growth of the magnetic field energy by a factor of $16^2$ would correspond to a magnetic Reynolds number of ${\mathcal R}_{\rm m} \sim 10^4$ (for \citealt{Pjanka:2020aa}, and less for \citealt{Oyang:2021aa}). This implies that the numerical magnetic diffusivities, $\eta_N$, of these codes, compared to the values physically expected in such discs, $\eta$, are too high by factors of at least 6 orders of magnitude. 
 
\cite{Pjanka:2020aa} provide a brief, preliminary discussion of the difficulty of accessing the physical parameter space using current numerical techniques, limited by current numerical processing power. They conclude that nevertheless ``the global structure and behavior of these models should be reflected properly''.  This is hard to square with the large discrepancy between the models and the reality of the fundamental measured disc parameter, $\alpha$. In contrast, \cite{Oyang:2021aa} note that the size of viscosity produced by MHD processes in their simulation is too small to be able to account for the superhump phenomenon that they are investigating. It is only once they artificially increase the viscosity by about an order of magnitude that they are able to account for the phenomenon.

Furthermore, the fact that the observed values of the viscosity parameter imply values of the plasma parameter $\beta \sim 1$ means that magnetic buoyancy effects may play a far greater role than is found in the simulations \citep{Tout:1992aa}. \cite{Balbus:1998aa} noted that the simulations vastly overestimate the scale at which resistive losses occur. We agree with this assessment. In view of all this, we suggest that the large discrepancy mentioned above may well imply that the nature of the dynamos found in the simulations is fundamentally different from that which occurs in real accretion discs. 

We have speculated that in the global numerical simulations it may be the large numerical diffusivities in those simulations that present the problem. If that is the case, it will be important to discover how close the numerical magnetic Reynolds number needs to be to the physical value of ${\mathcal R}_{\rm m} \approx  10^{10}$ in order that global disc dynamo simulations can achieve the required values of $\beta \sim 1$. Given the current limitations on computing power it may be that expecting to be able to compute realistic dynamo action in observable accretion discs using numerical MHD codes is, for the time being, a step too far.

{\bf Acknowledgments}. We thank both Referees and the Editor for useful comments and discussion.

{\bf Funding}. This work was supported by the Science and Technology Facilities Council (C.J.N., grant number ST/Y000544/1); and the Leverhulme Trust (C.J.N., grant number RPG-2021-380).

{\bf Declaration of Interests}. The authors report no conflict of interest.

{\bf Author ORCID}. C. J. Nixon, \url{https://orcid.org/0000-0002-2137-4146}; C. C. T. Pringle, \url{https://orcid.org/0000-0001-9908-626X}; J. E. Pringle, \url{https://orcid.org/0000-0002-1465-4780}.

\appendix
\section{Solution for arbitrary initial field structure}\label{appA}
As in the main text, we are considering a flow in the $xy$--plane with an inexorable linear shear flow of the form ${\bf u} = (U^\prime y, 0)$, with $U^\prime$ a constant. We assume the fluid to be incompressible, and to obey the standard MHD equations with a magnetic diffusivity $\eta$, so that the evolution equation for the magnetic flux function $A(x,y, t)$ is
\begin{equation}
\frac{\partial A}{ \partial t} + U^\prime y \frac{\partial A}{\partial x} = \eta \nabla^2 A\,. \label{Aevolve_app}
\end{equation}

We consider the evolution of $A$ on the domain $L_x \times L_y$. At time $t=0$, the flux function $A(x,y,0)$ can be expanded as a Fourier series in the form
\begin{equation}
A(x,y,0) = \sum_n \sum_k a_{n,k}  \exp (i \alpha n y) \exp (i \beta k x)\,,
\end{equation}
where $\alpha = 2 \pi / L_y$ and $\beta = 2 \pi / L_x$.

At later times we expect the form of $A$ to be
\begin{equation}
A(x,y,t) =  \sum_n \sum_k a_{n,k}(t) \exp (i \alpha n y) \exp (i \beta k[ x- U'ty])\,.
\end{equation}

Substituting this into Equation~\ref{Aevolve_app}, and using the fact that the modes evolve independently, we find that this can be satisfied provided that the coefficients $a_{n,k}(t)$ obey the equations
\begin{equation}
\frac{d a_{n,k}}{dt} = - \eta [ (\alpha n - \beta k U't)^2 + \beta^2 k^2]\: a_{n,k}\,.
\end{equation}

Thus we find that
\begin{equation}
a_{n,k}(t) = c_{n,k} \exp \tau_{n,k}(t)
\end{equation}
where, for $k \ne 0$ we have,
\begin{equation}
 \tau_{n,k}(t) =  \eta \left[ \frac{1}{3 \beta k U'} (\alpha n - \beta k U't)^3 - \beta^2 k^2 t \right]\,,
\end{equation}
and $c_{n,k}$ are constants.

For the $k = 0$ mode (for which the contribution to $\mathbf{B}$ is independent of $x$), we find that
\begin{equation}
a_{n,0} = c_{n,0} \exp [ - \eta \alpha^2 n^2 t ]\,.
\end{equation}

Note that in general the $c_{n,k}$ can take any values, provided that $c_{n,k} = c^*_{-n,-k}$, where the $^*$ denotes the complex conjugate.

The full solution for $A(x,y,t)$ is therefore
\begin{equation}
A(x,y,t) =  \sum_n \sum_k c_{n,k} \exp(-\tau_{n,k}(t)) \exp (i \alpha n y) \exp (i \beta k[ x- U'ty])\,,
\end{equation}
where
\begin{equation}
\tau_{n,k}(t) =  \begin{cases} 
	\eta \alpha^2 n^2 t & {\rm if}~k = 0 \\
	\eta \left[ \frac{1}{3 \beta k U'} (\alpha n - \beta k U't)^3 - \beta^2 k^2 t \right] & {\rm if}~k\ne 0\,.
   \end{cases}
\end{equation}

Thus for the particular case $A(x,y,t=0) = \sin (\beta x)$ we take the non-zero values of $c_{n,k}$ to be
\begin{equation}
c_{n, k} = \left\{ \begin{array}{ll}
                          -(1/2)i & n=0,k=1 \\
                          (1/2)i & n=0, k=-1 \\
                          0 & \mbox{otherwise}
                          \end{array}
                          \right.
\end{equation}

And for the case $A(x,y,t=0) = \sin (\beta x) \sin (\alpha y)$, we take
\begin{equation}
c_{n,k} = \left\{ \begin{array}{ll}
                         -(1/4) \exp[-\tau_{n,k}(0)] & (n,k) = \pm(1,1)\\
                         (1/4) \exp[-\tau_{n,k}(0)] & (n,k) = \pm(1,-1) \\
                         0 & \mbox{otherwise}
                         \end{array}
                         \right.
\end{equation}

\bigskip

The magnetic field is ${\bf B} = (\partial_y A, - \partial_x A)$. Thus
\begin{equation}
{\bf B} = \sum_n \sum_k ( i \alpha n - \ \beta k U't, - i \beta k) \: c_{n,k} \exp (i \alpha n y) \exp (i \beta k [x - U'ty]) \exp [ \tau_{n,k}(t)]\,.
\end{equation}

Then with the spatially averaged magnetic energy defined to be
\begin{equation}
\langle {\bf B}(t)^2 \rangle = \int_0^{2 \pi/\alpha} \int_0^{2 \pi/\beta} | {\bf B}(t) |^2 \, dx \, dy\,,
\end{equation}
we can use Parseval's Theorem to deduce that
\begin{equation}
\frac{ \langle {\bf B}(t)^2 \rangle} { \langle {\bf B}(0)^2 \rangle} = \frac { \sum_n \sum_k [ (\alpha n + \beta k U't)^2 + \beta^2 k^2] |c_{n,k} |^2 \exp [2 \tau_{n,k} (t) ] } { \sum_n \sum_k [ (\alpha^2 n^2 + \beta^2 k^2] |c_{n,k} |^2 \exp [2 \tau_{n,k} (0) ] }\,.
\end{equation}

\bibliographystyle{jpp}
\bibliography{nixon}
\end{document}